\documentclass[aps,prl,twocolumn,showpacs,epsfig]{revtex4}

\usepackage{graphicx}

\begin{document}

\title{A propensity criterion for networking  \\
 in an array  of coupled chaotic systems}
\author{F.T. Arecchi $^\clubsuit$$^\spadesuit$ \footnote{e-mail:arecchi@ino.it},
E. Allaria$^\spadesuit$ \footnote{e-mail:allaria@ino.it},
I. Leyva$^\diamondsuit$$^\spadesuit$\footnote{e-mail:ileyva@escet.urjc.es}}
\affiliation  {$^\clubsuit$   Dept.  of
Physics,  University  of   Florence,  Florence,  Italy\\
$^\spadesuit$ Istituto  Nazionale di  Ottica  Applicata, Largo  E.  Fermi, 6  I50125
Florence,    Italy\\
$^\diamondsuit$    Universidad   Rey    Juan
Carlos. c/Tulipan s/n. 28933 Mostoles Madrid, Spain}

\date{\today}

\begin{abstract}

We  examine the mutual
synchronization of a one dimensional  chain of chaotic identical objects in the
presence  of a  stimulus applied  to the  first site.  
We first describe the characteristics of the local elements, 
and then the process whereby a global nontrivial behaviour emerges. 
A propensity criterion for networking is introduced, consisting in the coexistence within the attractor
of a localized chaotic region, which displays high sensitivity to external stimuli,
and an island of stability, which provides a reliable coupling signal to the neighbors in the chain.
Based on this criterion we compare homoclinic chaos, recently explored in lasers and conjectured
to be typical of a single neuron, with Lorenz chaos.

\end{abstract}

\maketitle

An open problem in science is how to build a semantic network on minimal 
assumptions. 
In a living brain 
an external stimulus localized at some 
input spreads over a large assembly of coupled neurons
building up a collective state univocally corresponding to the stimulus.
The current conjectures and the preliminary experimental 
evidence \cite {rieke, singer} on this time dependent networking 
problem lead to a new paradigm wherein perceptions require mutual
synchronization of neuronal spikes. 
A model of dynamical encoding by networks of competing neurons has been 
recently introduced \cite{rabinovichPRL01}, however the core issue of 
synchronizing a large number of neurons as it appears from experiments 
\cite {singer} has not been addressed thus far.

With reference to this issue we introduce a minimal model built upon 
simplicity requirements, namely
we consider linear and symmetric interneuron coupling; furthermore, we 
take just nearest neighbor coupling, avoiding architecturally complicated  
connections. Under these assumptions we address the question of how 
complexity arises, calling  
complexity the fact that a composite  system
displays  collective properties not  directly deducible from the dynamical
behavior of its constituent elements \cite{anderson}.

Here we focus  on the  collective response  of an  array  of coupled
dynamical  objects to  a localized  external stimulus,  and  provide a
propensity  criterion for  networking, that is, for  organizing in  a
collective state univocally related  to the stimulus.
Having in mind time  dependent networking, as it is the case of
biological  communication,  we  examine  mutual
synchronization of a one dimensional  chain of identical objects in the
presence  of a  stimulus applied  to the  first site.  
We first describe the characteristics of the local elements, 
and then the process whereby a global nontrivial behaviour emerges. 

Two independent requirements  must be fulfilled  by each  element.
First, it should have  a sensitivity  region in order  to easily respond  to the
neighbour coupling; second it has to provide a strong enough signal to
relay the input stimulus along  the whole array. The first requirement
suggests to recur to dissipative  chaotic systems; in fact modifying a
regular individual dynamics confined to a stable attractor would imply
a consistent expenditure of energy and time \cite{amit}, whereas sweeping
through  the manifold  of  unstable periodic orbits which make up a
chaotic attractor is a fast cost-less operation \cite{auerbach}.
The  second  requirement is  conflicting  with  the  previous one; in fact, it
implies islands of stability within the chaotic orbit, out of which to
extract a reliable driver for the next neighbour.

The twofold problem is solved recurring to large spikes emerging out of
a small chaotic background; indeed a weak  inter-site 
coupling  will  provide  a  discrete
synchronization  associated  only  with  the large  spikes, the
chaotic background being not effective. This
discrimination will amount to the approximation inducing the 
transition from an individual to a collective description.
At variance with the standard chaotic
synchronization scheme \cite{PecoraCarrol} where two identical systems
synchronize   along   the    whole   orbital   evolution,   here   the
synchronization occurs because a large  spike of one system is forcing
the neighbour to escape away from its chaotic region, thus yielding 
its own spike \cite{Leyva_uni,Leyva_bidi}.

Among  the chaotic  systems,  those  best  suited  to  the
emergence of a new hierarchical  level should thus be characterized by
temporal  windows  of  stability  and  chaos within  each  orbit. 

We  give  substance to  these  considerations  with  reference to 
the heteroclinic transfer  forth and back between a  saddle focus (SF)
and  a  saddle node  (SN)  under  the  so called  Shilnikov  condition
\cite{Shilnikov}.    Such   a   behavior   has  been   explored   both
experimentally \cite{arecchi_87} and theoretically \cite{pisarchik_00}
with reference to a CO$_2$  laser with feedback.  For sake of brevity,
it will be called HC (homoclinic chaos) \cite{allaria}.
In fact the mere homoclinic return to
SF would provide a chaotic transient \cite{dhamale99}
but not assure a regolar motion away from SF
; on the contrary, the further presence of SN 
yields the stability island necessary for networking.

The dynamics is characterized by a sequence of
spikes with widely fluctuating time intervals T. Such a structure
underlies spiking behavior in many neuron \cite{hh,feudel},
chemical \cite{argoul}, laser \cite{arecchi_87}, and El Ni\~no \cite{timmermann} systems. It is
important to note that this dynamics is highly nonuniform, in
the sense that the sensitivity to small perturbations is high
only in the vicinity of S along the unstable directions. A
weak noise thus may influence T significantly \cite{noise}.  .

In the system we consider \cite{pisarchik_00} the chaotic behavior is
confined in a small neighborhood of SF where it fulfills the Shilnikov
condition \cite{Shilnikov} $ \gamma <  \alpha$ ,$\gamma$  being the real part of the
expansion rate  on the unstable manifold  of SF and  $\alpha$ the 
contraction  rate   on  the  stable  manifold   of  SF.  
Precisely the linearized dynamics around SF is
ruled  by  the  leading eigenvalues  $(-\alpha, \gamma  \pm i \omega)$ with $\alpha = 7.58$, $\gamma = 3.52$,
all the other eigenvalues having very large negative real parts.
The exiting  trajectories along the
unstable manifold reenter the stable  one after a large orbit in phase
space, corresponding  to the heteroclinic  approach to SN.   The phase
space orbit appears as  a confined chaotic tangle, concentrated around
the saddle focus which is connected to a wide regular section, the two parts 
closing the orbit (Fig. 1-left).

For  such a system  synchronization has  been demonstrated  against an
external  clock \cite{allaria},  or against  a previous  time  slot of
itself, presented  after a  suitable delay \cite{DSS}  or after  a low
pass filter \cite{Bursting}. Furthermore, HC is robust against noise
\cite{noise}.  As we go from one system to an array of coupled identical systems 
\cite{Leyva_uni,Leyva_bidi}, mutual synchronization occurs either spontaneously 
or as a response to an external forcing applied to a single site.

We recall \cite{Leyva_bidi} that the coupling is realized 
by replacing in one of the HC equations \cite{pisarchik_00} a scalar component
$x_1^i$ ($i=$ site index) with 
$x_1^i [1+\epsilon(x_1^{i+1}+ x_1^{i-1}-2 <x_1^i>)]$
where $<x_1^i>$ denotes a moving time average. The range of coupling strengths
$\epsilon$ considered here is between $0$ and $0.25$.

To show the advantage of HC in comparison to more conventional types of 
chaotic behaviour, we consider the well known Lorenz model \cite{lorenz}.
In terms  of fixed points,  also the rectified  Lorenz system(i.e.,the
Lorenz  model  plus  an  inversion  operation around  the  origin)  is
characterized by one SF and one SN; however for the standard values of
the  control   parameters b=8/3,$\sigma$=10, r=28,   the
eigenvalues  around SF  are $(-\alpha  \pm  i  \omega,  \gamma)$ with
$\alpha$=14.48  and  $\gamma=0.4119$, thus
very far away from the Shilnikov condition.

\begin{figure}
\begin{center}
\includegraphics[width=9cm]{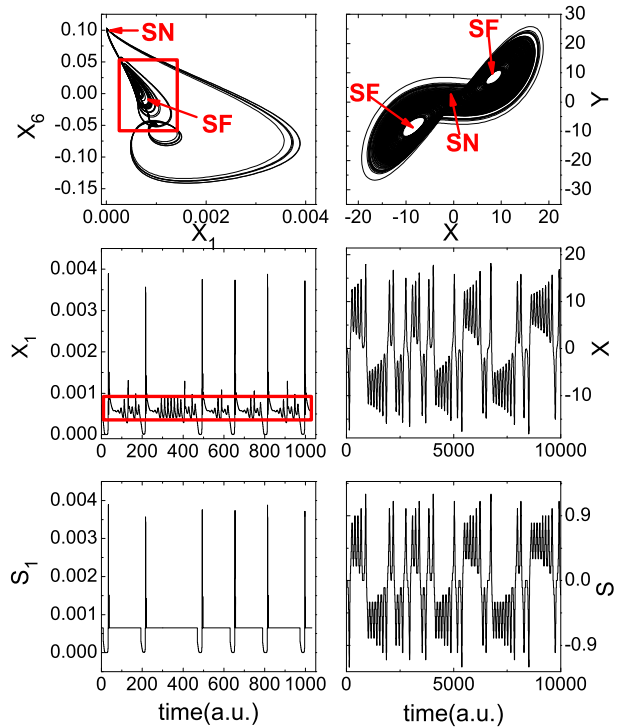}
\caption{
Comparison between HC(homoclinic chaos) \cite{arecchi_87} (left column)
and Lorenz chaos (right column). The top row is a phase space projection
over two dynamical variables; SF  denotes the saddle focus and SN the saddle
node, on the left the two SF map one onto the other after an inversion 
($x \rightarrow -x$, $y \rightarrow -y$) around the origin.
The intermediate row shows the time series for variables $x_1$ of  HC (it
represents the laser intensity in the case of the CO$_2$ laser) and  $x$ of
Lorenz; in the former case, a suitable threshold cuts off the chaotic
background; in the latter case, no convenient region for thresholding can be isolated.
Bottom row, left :  after threshold, the new variable $S (t)$ alternates spikes
with flat regions where the systems has a high sensitivity and short
refractive windows where the intensity $S (t)$ goes to zero.
}
\end{center}
\end{figure}

In fact,  morphologically the two  cases appear very  different; thus,
even  though   a  homeomorphism  should   map  one  over   the  other,
introduction  of extra  operations, such  as a  sensitivity threshold,
renders such a transfer impossible, as one can see by comparing left and right
columns of Fig.~1 before and after thresholding (intermediate and bottom rows).

\begin{figure}
\begin{center}
\includegraphics[width=9cm]{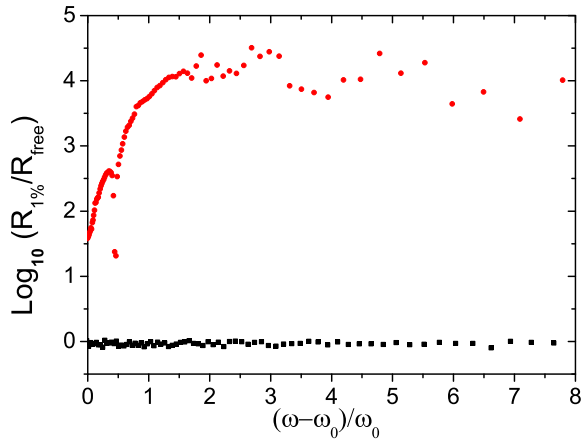}
\caption{
Coherence parameter $R_{1\%}
$ for a driving signal consisting of a 1 \%
periodic perturbation of a control parameter normalized to $R_{free}$ (
the $R$ value
in the absence of perturbation), plotted versus the normalized distance of the
perturbation frequency from the natural value $\omega _0$.
Circles : HC; squares : Lorenz.
For HC the ratio $ \frac {R_ {1 \%}} {R_{free}} 
$ is about 30 
at $\omega _0$
and it increases up to $10^4$ for $\omega$ twice or more the natural frequency;
indeed, since synchronization means forcing HC 
away from the SF region, frequencies higher than 
the natural one are easier to synchronize, whereas for smaller frequencies 
HC may have a spontaneous escape from SF.
To smooth the plot, a
small amount of noise (0.5 \%) has been added.
For the Lorenz case, the indicator is always at 1.
}
\end{center}
\end{figure}

Precisely, the  time plot  of one of  the HC variables 
consists of  a train of  identical spikes, separated  by a
variable interspike interval ($ISI$)  filled with a rather small chaotic
signal confined within a stripe thinner than $1/5$ of the spike height.
As we assemble a large number of such dynamical systems in a network 
and identify an optimal mutual coupling, which yields an efficient transfer 
of the input information along the network, a reduction by a factor $3$ of that coupling 
makes already the transfer inefficient, as shown later in Fig.~3; thus a natural thresholding 
is provided by the alternation large-small which makes the bottom plot on Fig.~1~left equivalent to the 
intermediate one for networking purposes, without need to apply the threshold as an artifact.

\begin{figure}
\begin{center}
\includegraphics[width=9cm]{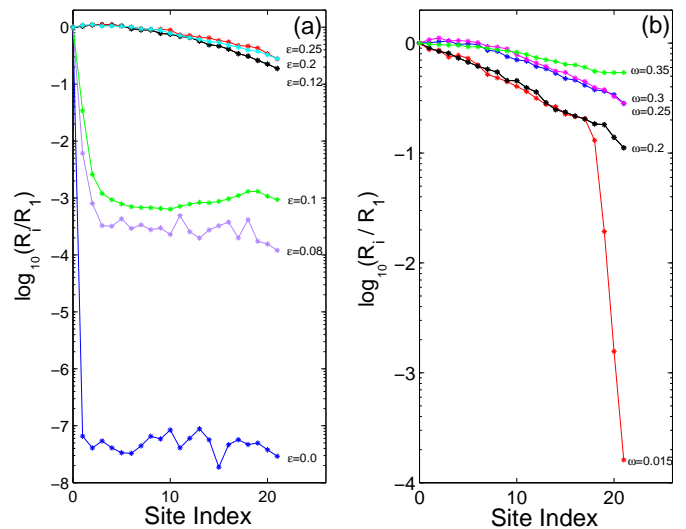}
\caption{
Propagation along a chain of $N=20$ sites of the coherence value 
$R_1$ of the first site exposed to an external stimulus. 
a) For a fixed input frequency close to the natural one $\omega_0$
and different coupling strengths $\epsilon$. 
b) for $\epsilon = 0.2$ and different frequency ratios $\omega / \omega_0$.
}
\end{center}
\end{figure}

The emerging
spikes display  a chaotic time  occurrence,which is epitomized  by the
correlation properties of $ISIs$. Mathematically, this thresholded chaos
is described  as $S(t)=S_o \sum_l  \delta (t -\tau_l)$  where $\tau_l$
are the  chaotic spike positions  and ($ISI$)$_l=\tau_l-\tau_{l-1}$.  In
presence of an external driving  signal, the spikes can synchronize to
it  \cite{allaria}; in  case  of  many coupled  systems  of this  type
located on an array, wide  parameter ranges are found within which the
individual sites  mutually synchronize their  own spikes, so  that the
space-time  plot of  the spike  positions at  each site  appears  as a
regular fabric  \cite{Leyva_uni,Leyva_bidi}.

Such an  easy mutual  synchronization as the  response to  a localized
input represents a semantic property absent in the Lorenz
case  \cite{PecoraCarrol} where  there is  no apparent  scale separation
where to consider a threshold.
We thus identify the  propensity criterion  with the  presence within
each orbit of  two very different amplitude scales,  that of the large
spikes and that  of the small chaotic background,  the mutual coupling
between  sites being  operated by  the spikes  whereas  the background
represents the high sensitivity region within which the coupling takes
place.

As  indicator  of successful  networking  we  take the  coherence
parameter \cite{boccaletti_02}
\begin{equation}
R=\frac{<ISI>}{\delta ISI}
\end{equation}
where  $\delta$ISI is  the square  root of  the $ISI$  variance ($\delta
ISI=\sqrt{<(ISI-<ISI>)^2>}$).  $R$  has been used to measure
the amount of synchronization of a single system
to a periodic stimulus;
when  synchronization propagates along  a chain,  $R$ can  be measured
anywhere since it has almost the same value on all sites.

Notice that standard chaotic synchronization can propagate along a
chain of generic chaotic systems, however,
in general there is no propensity of the first site to synchronize
to a weak external input,  as shown by the corresponding indicator $R$
(Fig.~2) for the systems of Fig.~1,
either in the case of propensity (left) and no-propensity. 
The stimulus consists of a sinusoidal perturbation of frequency $\omega$ 
applied to a control parameter in the first site of the chain, with an amplitude
of $1\%
$ of the unperturbed parameter value.
The $R$ value is about $30$ for $\omega$ at 
the natural frequency of the system $\omega _0 =2 \pi/<ISI>$, and increases
up to $10^4$ for larger frequencies, whereas for the Lorenz case it is consistentely 
$R=1$ (no coherence).
We now address the crucial question how the coherence $R_1$
induced on the first site $i=1$ propagates along the array, for different
coupling strengths $\epsilon$ and frequencies $\omega $ of the input signal.
As shown in Fig.~3, reducing the coupling from $\epsilon=0.25$ to $\epsilon=0.08$
reduces $R$ by three orders of magnitude, thus showing the natural thresholding effect 
occurring in a network of HC systems, without having to explicitely take care
for the operation leading from Fig.1(left-middle) to Fig.1(left-bottom).

In conclusion, we have introduced the notion of semantic network as an
array of coupled identical chaotic systems which assume a collective
state in presence of a localized periodic stimulus.
The propensity criterion appears morphologically as a confinement of the
chaotic tangle within a small region of the total attractor; the
corresponding indicator is a tremendous increase in the $R$ parameter near
the natural frequency.

At variance with complete synchronization \cite{PecoraCarrol}, here we consider only
synchronization of large spikes, intercalated by intervals not effective
in acting over the neighbours,yet highly sensitive to external signals.
This split of the dynamical orbit into two different regions is the condition
to build a nontrivial collective state as response to a localized stimulus.

Two further relevants issues will be dealt with elsewhere, namely 
i)how the network organizes in presence of two stimuli of different period 
applied to opposite sites; and  ii)learning of a complex pattern, coded by a 
non periodic sequence of spikes limited in time.

\vspace{1.5 cm}
Authors are indebted to S. Boccaletti and P. Lauro-Grotto for useful discussions.
E.A. and I.L. acknowledge support from the European Contract CoSyC of SENS
No. HPRN-CT-2000-00158.


\begin{thebibliography}{99}

\bibitem{rieke} F. Rieke {\sl et al.} {\it "Spikes : Exploring the Neural Code"}, MIT Press, Cambrige, MA, (1997).

\bibitem{singer} W. Singer and C.M. Gray, 
Annu. Rev. Neurosci. {\bf 18}, 555 (1995).

\bibitem{rabinovichPRL01} M. Rabinovich, A. Volkovskii, P. Lecanda, R. Huerta, H. D. I. Abarbanel, and G. Laurent
Phys. Rev. Lett {\bf 87}, 068102 (2001).

\bibitem{anderson} P.W. Anderson, 
Science {\bf 177} , 393 (1972).

\bibitem{amit} D.J. Amit  {\it Modeling Brain Function :  The World of
Attractor  Neural  Networks}  Cambridge  University  Press;  (November
1989).

\bibitem{auerbach}   D.  Auerbach,   P.   Cvitanovic,  J.P.   Eckmann,
G.  Gunaratne  and  I.  Procaccia,
Phys. Rev. Lett. {\bf 58}, 2387-2389 (1987).

\bibitem{PecoraCarrol}     L.M.    Pecora     and     T.L.    Carroll,
Phys. Rev. Lett. {\bf 64}, 821, (1990).

\bibitem{Leyva_uni}   I.  Leyva,   E.  Allaria,   S.   Boccaletti  and
F. T. Arecchi, 
(e-print: nlin.PS/0210042).

\bibitem{Leyva_bidi}   I.  Leyva,  E.   Allaria,  S.   Boccaletti  and
F.   T.   Arecchi,
(e-print:nlin.PS/0302008).

\bibitem{Shilnikov}  L. P. Shilnikov, 
Sov. Math.  Dokl. {\bf  6}, 163 (1965);
L. P. Shilnikov, Int. J. Bif. Chaos {\bf 4}, 489 (1994).

\bibitem{arecchi_87} F.T.  Arecchi, R.  Meucci and W.  Gadomski,
Phys. Rev. Lett.,  {\bf 58}, 2205 (1987).

\bibitem{pisarchik_00}  A.N. Pisarchik,  R. Meucci  and  F.T. Arecchi,
Eur. Phys. J. D {\bf 13}, 385 (2001).

\bibitem{allaria}  E.  Allaria,  F.   T.  Arecchi,  A.  Di  Garbo  and R. Meucci,
Phys. Rev. Lett. {\bf 86}, 791 (2001);

\bibitem{dhamale99} M.  Dhamala and Y.C. Lai, 
Phys. Rev. E {\bf 59}, 1646 (1999).

\bibitem{hh} A.L. Hodgkin and A.F. Huxley, J. Physiol. Londo {\bf 117}, 500 (1952); 
E.M. Izhikevich, Int. J. Bifurcation Chaos {\bf 10}, 1171 (2000).

\bibitem{feudel}
U. Feudel {\it et al.}, Chaos {\bf 10}, 231 (2000).

\bibitem{argoul}
F. Argoul, A. Arne\'odo, and P. Richetti, Phys. Lett. {\bf 120A}, 269
(1987).

\bibitem{timmermann}
A. Timmermann and F.F. Jin, Geophys. Res. Lett. {\bf 29}, 10.1029/
2001GLO13369 (2002).

\bibitem{noise}  C.S.  Zhou,  J.  Kurths, E.  Allaria,  S.  Boccaletti,
R. Meucci and F.T. Arecchi, 
Phys. Rev. E {\bf 67}, 15205(R) (2003).

\bibitem{DSS} F.T.  Arecchi, R.  Meucci, E. Allaria,  A. Di  Garbo and
L.S. Tsimring,
Phys. Rev. E {\bf 65}, 046237  (2002).

\bibitem{Bursting}   R.  Meucci,   A.   Di  Garbo,   E.  Allaria   and
F. T. Arecchi, 
Phys. Rev. Lett.  {\bf 88}, 144101 (2002).

\bibitem{lorenz} E.N. Lorenz,
J. Atmos. Sci. {\bf 20}, 
130 (1963). 

\bibitem{boccaletti_02} S. Boccaletti, J. Kurths, G. Osipov, D.L. Valladares and C.S. Zhou,
Phys. Reports {\bf 366}, 1-101 (2002). 



\end{thebibliography}
\end{document}